\documentclass[epj]{webofc}
\usepackage[utf8]{inputenc}
\usepackage[varg]{txfonts}   % Web of Conferences font
\usepackage{booktabs}
\usepackage{xcolor}
\definecolor{darkred}{rgb}{0.4,0.0,0.0}
\definecolor{darkgreen}{rgb}{0.0,0.4,0.0}
\definecolor{darkblue}{rgb}{0.0,0.0,0.4}
\usepackage[bookmarks,linktocpage,colorlinks,
    linkcolor = darkred,
    urlcolor  = darkblue,
    citecolor = darkgreen]{hyperref}
\usepackage{alltt}
\wocname{EPJ Web of Conferences}
\woctitle{Lattice2017}
\newcommand{\QED}{\mathrm{QED}}
\newcommand{\Wilson}{\mathrm{Wilson}}
\newcommand{\csw}{c_{\mathrm SW}}

\newlength{\dWidth}     % digit width
\newcommand{\D}{\makebox[\dWidth]{}}
\begin{document}
%%%%%%%%%%%%%%%%%%%%%%%%%%%%%%%%%%%%%%%%%%%%%%%%%%%%%%%%%%%%%%%%%%%%%%%%%%%%%
%
\selectlanguage{english}
%----------------------------------------------------------------------------
\title{%
An update on the BQCD Hybrid Monte Carlo program
}
%----------------------------------------------------------------------------
\author{%
\firstname{Taylor Ryan} \lastname{Haar}\inst{1} \and
\firstname{Yoshifumi} \lastname{Nakamura}\inst{2} \and
\firstname{Hinnerk}  \lastname{Stüben}\inst{3}\fnsep\thanks{Speaker, \email{hinnerk.stueben@uni-hamburg.de}}
% etc.
}
%----------------------------------------------------------------------------
\institute{%
CSSM, Department of Physics, The University of Adelaide, Adelaide, SA, Australia 5005
\and
RIKEN Advanced Institute for Computational Science, Kobe, Hyogo 650-0047, Japan
\and
Universität Hamburg, Regionales Rechenzentrum, 20146 Hamburg, Germany
}
%----------------------------------------------------------------------------
\abstract{% 
We present an update of BQCD, our Hybrid Monte Carlo program for
simulating lattice QCD.  BQCD is one of the main production codes of the
QCDSF collaboration and is used by CSSM and in some Japanese finite
temperature and finite density projects.  Since the first publication of
the code at Lattice 2010 the program has been extended in various ways.
New features of the code include: dynamical QED, action modification in
order to compute matrix elements by using Feynman-Hellman theory, more
trace measurements (like ${\mathrm Tr}(D^{-n})$ for $\kappa$,
$c_{\mathrm{SW}}$ and chemical potential reweighting), a more flexible
integration scheme, polynomial filtering, term-splitting for RHMC, and a
portable implementation of performance critical parts employing SIMD.
}
%----------------------------------------------------------------------------
\maketitle
%----------------------------------------------------------------------------
\section{Introduction}

BQCD is a Hybrid Monte Carlo program for simulating lattice QCD with
dynamical Wilson fermions.  It was first published at Lattice 2010
\cite{Nakamura:2010qh} and has been used by several groups: the
QCDSF-UKQCD collaboration \cite{Stuben:2000wm, Gockeler:2007rm,
  Cundy:2009yy, Horsley:2015vla, Chambers:2014qaa, Chambers:2015bka},
CSSM \cite{Chambers:2014qaa, Chambers:2015bka, Haar:2016bwe,
  Hollitt:2017tkf}, Japanese finite density \cite{Jin:2013wta,
  Jin:2015taa} and finite temperature \cite{Jin:2017jjp} projects, and
the RQCD collaboration \cite{Bali:2017pdv}.

Here we report on extensions and optimizations that were made meanwhile
and give an update on compute performance.  The code and a manual can be
downloaded from \cite{bqcd-download}.  New features of the program are:

\begin{itemize}
\item \emph{Actions:} hopping term with chemical potential, clover
  $O(a)$ improved Wilson action plus a CPT breaking term, QCD+QED,
  QCD+Axion. See section~\ref{sec-actions}.
\item \emph{Algorithms:} polynomial filtering, a
  generalized multiscale integration scheme, truncated RHMC, Zolotarev
  approximation. See section~\ref{sec-algorithms}.
\item \emph{Compute performance optimizations}: explicit vectorization
  with SIMD intrinsics, improvement of MPI communication.  See
  section~\ref{sec-opti}.
\end{itemize}

%----------------------------------------------------------------------------
\section{Actions} \label{sec-actions}

\subsection{Gauge actions}

Implemented are the Wilson gauge action and an improved gauge
action, see~\cite{Nakamura:2010qh}.

\subsection{Fermion actions}

At the time of \cite{Nakamura:2010qh} the program could simulate the
standard Wilson fermion action $S^\Wilson_F$, $O(a)$ clover improved
Wilson fermions, and the SLiNC fermion action \cite{Cundy:2009yy}.  The
new version can also simulate:

\begin{itemize}
\item the hopping term with chemical potential $\mu$ \cite{Jin:2013wta,Jin:2015taa}
\begin{equation}
\begin{split}
S_F = \sum_x
\Bigg\{
\bar{\psi}(x) \psi(x) 
&- \kappa \sum_i^3 \left[
\bar{\psi}(x) U_i^\dagger(x-\hat{i})(1+\gamma_i) \psi(x-\hat{i})+\bar{\psi}(x) U_i(x)(1-\gamma_i) \psi(x+\hat{i})
\right] \\
&- \kappa  \left[
\bar{\psi}(x) U_4^\dagger(x-\hat{4})(1+\gamma_4)e^{-\mu} \psi(x-\hat{4})+\bar{\psi}(x) U_4(x)(1-\gamma_4)e^\mu \psi(x+\hat{4})
\right]
\Bigg\}
\end{split}
\end{equation}
\item the clover $O(a)$ improved Wilson action plus a CPT breaking term with
  coefficient $\lambda$ and a $4 \times 4$ matrix $H$
  \cite{Chambers:2014qaa,Chambers:2015bka}
\begin{equation}
S_F = S^\Wilson_F - \frac{i}{2} \kappa\,\csw 
\sum_x \big[ \bar{\psi}(x) \sigma_{\mu\nu} F_{\mu\nu}(x) \psi(x) + \kappa\,\lambda \bar{\psi}(x) H \psi(x) \big]
\end{equation}
\end{itemize}

\subsection{QCD+QED}

The program can simulate QCD+QED \cite{Horsley:2015vla} using the action
\begin{equation}
S = S_G + S_A + \sum_q S_F^q \;.
\end{equation}
$S_G$ is an SU(3) gauge action, $S_A$ is the non-compact U(1) gauge
action
\begin{equation}
S_A = \frac{\beta_{\QED}}{2} \sum_{x, \mu < \nu}
\left[ A_\mu(x) + A_\nu(x + \hat\mu) - A_\mu(x + \hat\nu) - A_\nu(x) \right]^2 \;,
\end{equation}
and the fermion action for flavour $q$ is
\begin{equation}
\begin{split}
S_F^q 
&= \sum_x {\Bigg\{} \kappa_q \sum_\mu \left[ \overline{q}(x) (\gamma_\mu -1) e^{-i Q_q A_\mu(x) } \tilde {U}_\mu(x) q( x + \hat\mu) \right.
\\
&- \left. \overline{q}(x) (\gamma_\mu +1) e^{ i Q_q A_\mu(x-\hat\mu) } \tilde {U}^\dagger_\mu(x-\hat\mu) q(x-\hat\mu) \right]
 +\;  \overline{q}(x) q(x)  -\frac{1}{2} \kappa_q \csw \sum_{\mu,\nu} \overline{q}(x) \sigma_{\mu \nu} F_{\mu \nu}(x) q(x) \Bigg\} \;,
\end{split}
\end{equation}
where $Q_u = +2/3$, $Q_d = Q_s = -1/3$ and $\tilde {U}_\mu$ is a singly
iterated stout link.

\subsection{QCD+Axion}

The program can simulate QCD+Axion \cite{thiscontrib235} using the action
\begin{equation}
S = S_G + S_a + \sum_q S_F^q \;.
\end{equation}
$S_a$ is scalar action for the axion field  $\phi_a$
\begin{equation}
S_a = \kappa_a \sum_x \sum_\mu \Big (\phi_a(x) - \phi_a(x+\mu) \Big)\phi_a(x) \,, \\
\end{equation}
and the fermion action for flavour $q$ in the case of Wilson fermions is 
\begin{equation}
\begin{split}
S_F^q &=  \sum_{x} \bar{q}(x)   
\Big[1 +  (\kappa_q  \lambda_q + f_{\mathrm{inv}} \phi_a )\gamma_5  \Big] q(x)  \\
 &- \kappa_q \sum_{x,\mu} \Big[
    \bar{q}(x) (1 - \gamma_\mu) U_\mu(x) q(x + a\hat{\mu})
  + \bar{q}(x - a\hat{\mu}) (1 + \gamma_\mu) U_\mu^\dagger (x- a\hat{\mu}) q(x) \Big]
 \,,
\end{split}
\end{equation}
where
\begin{equation}
\kappa_q = {1\over 2am_q +8}\,,\quad
\lambda_q = i2am_q {\theta \over N_f}\,,\quad
f_{\mathrm{inv}} = i2 \kappa_q  m_q   { \sqrt{\kappa_a }\over f_a N_f}\,.
\end{equation}

%----------------------------------------------------------------------------
\section{Measurements}

The following quantities can be measured with BQCD:
 plaquettes (quadratic and rectangular),
 topological charge (cooling method),
 Polyakov loop,
 Wilson flow,
 traces of the fermion matrix ($\mathrm{Tr}\,(M^{-1})$,
$\mathrm{Tr}\,(\gamma_5M^{-1})$,
$\mathrm{Tr}\, (M^{\dagger}M)^{-1}$),
 quark determinant with chemical potential,
 smallest and largest eigenvalue of the Dirac matrix,
 meson and baryon propagators.

%----------------------------------------------------------------------------
\section{Algorithms} \label{sec-algorithms}

%The following algorithms are implemented in BQCD: Rational Hybrid Monte
%Carlo, multiple time scales (with either nested or generalized
%integrators), Filtering methods (Hasenbusch, polynomial filtering), truncated RHMC, various solvers (\emph{cg}, BiCGstab, GMRES, GCRODR,
%multishift \emph{cg}, block multishift \emph{cg}).

In addition to nested integrators for multiple time scales, a generalized integration scheme \cite{Haar:2016bwe} has been implemented.
This is where separate integration schemes for each action term are superimposed onto a single time step evolution.
This allows the integration step-sizes for each action term to be completely independent of the others.

Rational Hybrid Monte Carlo (RHMC) is implemented with rational approximations from the Remez algorithm.
In the case of approximating $(W^\dag W)^{-1/2}$, an alternative rational function is available, namely the Zolotarev optimal rational approximation (see e.g. \cite{Chiu:2002eh} for an explanation).

Alongside Hasenbusch filtering, there are two new filtering methods, one of which applies exclusively to RHMC:
\begin{itemize}
	\item \emph{Polynomial filtering} applies to both RHMC and standard HMC fermions, and is the application of a polynomial filter $P(W^\dag W)$ to split the fermion action into several terms:
\begin{equation}
	S_{PF} = \overline{\psi}_1 P(W^\dag W) \psi_1 + \overline{\psi}_2 P(W^\dag W)^{-1} W^\dag W \psi_2.
\end{equation}

	\item \emph{Term-splitting for RHMC} splits the sum in the rational approximation $R(W^\dag W)$ for RHMC into several terms, giving action
\begin{equation}
	S_{tRHMC} = \overline{\psi}_1 R_{1,t}(W^\dag W) \psi_1 +  \overline{\psi}_2 R_{t+1,N}(W^\dag W) \psi_2
\end{equation}
where
\begin{equation}
	R_{i,j}(K) = c_n^{\delta_{i1}} \sum_{k=i}^j \frac{W^\dag W + a_k}{W^\dag W + b_k},
\end{equation}
$a_k, b_k$ are ordered decreasing.

\end{itemize}

BQCD has a wide range of iterative solvers: \emph{cg}, BiCGstab, GMRES, GCRODR, multishift \emph{cg}, block multishift \emph{cg}.

%----------------------------------------------------------------------------
\section{Optimization of compute performance} \label{sec-opti}

\subsection{SIMD vectorization and MPI}

In addition to parallelization with MPI and OpenMP a third level of
parallel implementation was introduced for solvers: SIMD vectorization
with SIMD intrinsic functions.  The SIMD implementation is generic,
i.e.\ it works for any size of SIMD vectors.  In order to achieve this,
the data layout of arrays had to be changed.  All arrays (for gauge,
spin-colour and clover fields) now have SIMD vectors as the smallest
structure.  In Fortran notation the gauge field is defined in the
following way
\begin{quote}
\begin{tabular}{ll}
old: & \verb|complex(8) :: u(3, 3, volume/2)|\\
new: & \verb|real(8) :: u(SIMDsize, re:im, 3, 3, volume/2 / SIMDsize)|\\
\end{tabular}
\end{quote}
and the new layout of the spin-color field is
\begin{quote}
\begin{tabular}{ll}
old: & \verb|complex(8) :: a(4, 3, volume/2)|\\
new: & \verb|real(8) :: a(SIMDsize, re:im, 2, 3, volume/2 / SIMDsize, 2)|\\
\end{tabular}
\end{quote}
where the $4$ spin components of the spin-colour field are split into $2
+ 2$ components which optimizes MPI communication in $t$-direction.  The
clover arrays, for which a packed format is used, were changed
accordingly.

At the single core level the SIMD code is about 2 times faster than the
corresponding Fortran code.  With this speed-up computations are
increasingly dominated by communication and improvement of MPI
communication becomes important.  Hence, the following MPI optimizations
where made:

\begin{itemize}
\item The overhead introduced by the reduction to two-component spinors
  was minimized. Previously the projection was done for the whole local
  volume, now it is only done for boundary sites, and there is no
  projection in the $t$-direction needed any more.
\item All MPI 'buffers' are consecutive in memory and aligned to SIMD
  vector boundaries.
\item Communication can overlap with computation. This is implemented
  with MPI plus OpenMP, where the \emph{master thread} communicates
  while the other threads compute.
\end{itemize}

In tables \ref{tab-cray} and \ref{tab-bgq} performance figures are
listed for machines and lattices that are currently used in production.
The optimized code runs between 1.3 and 1.7 times faster.

\begin{table}[thb]
\small \centering
\caption{Double precision performance of the \emph{cg}-solver of BQCD on
  a Cray XC40 (24 cores per node).}
\label{tab-cray}
\begin{tabular}{ccccccccc}\toprule
& \multicolumn{4}{c}{$48^3\times96$ lattice} 
& \multicolumn{4}{c}{$64^3\times96$ lattice}\\
\cmidrule(lr{0.5em}){2-5}
\cmidrule(lr{0.5em}){6-9} 
& \multicolumn{2}{c}{Fortran} & \multicolumn{2}{c}{SIMD}
& \multicolumn{2}{c}{Fortran} & \multicolumn{2}{c}{SIMD} \\
\cmidrule(lr{0.5em}){2-3}
\cmidrule(lr{0.5em}){4-5}
\cmidrule(lr{0.5em}){6-7}
\cmidrule(lr{0.5em}){8-9} 
        & per core & overall & per core & overall & per core & overall & per core & overall \\  
\#cores & Mflop/s  & Tflop/s & Mflop/s  & Tflop/s & Mflop/s  & Tflop/s & Mflop/s  & Tflop/s \\
\addlinespace
\D1536 &\D930 &\D1.4 & 1557 &\D2.4 & 785 &\D1.2 & 1028 &\D1.6 \\
\D3072 &\D949 &\D2.9 & 1516 &\D4.7 & 795 &\D2.4 & 1231 &\D3.8 \\
\D6144 & 1222 &\D7.5 & 1558 &\D9.6 & 876 &\D5.4 & 1419 &\D8.7 \\
\D9216 & 1253 & 11.5 & 1775 & 16.4 & --  & --   & --   & --   \\
 12288 & 1274 & 15.7 & 1678 & 20.6 & 927 & 11.4 & 1572 & 19.3 \\
\bottomrule
\end{tabular}
\end{table}

\begin{table}[thb]
\small
\centering
\caption{Double precision performance of the \emph{cg}-solver of BQCD on
  an IBM BlueGene/Q (8192 cores, a \emph{midplane}, is the smallest
  partition that has a fully wired torus network and with our SIMD
  implementation the largest possible partition for the $48^3\times96$
  lattice, where the lattice volume per core is $48\times3^3$).}
\label{tab-bgq}
\begin{tabular}{ccccccccc}\toprule
& \multicolumn{4}{c}{$48^3\times96$ lattice} 
& \multicolumn{4}{c}{$64^3\times96$ lattice}\\
\cmidrule(lr{0.5em}){2-5}
\cmidrule(lr{0.5em}){6-9} 
& \multicolumn{2}{c}{Fortran} & \multicolumn{2}{c}{SIMD}
& \multicolumn{2}{c}{Fortran} & \multicolumn{2}{c}{SIMD} \\
\cmidrule(lr{0.5em}){2-3}
\cmidrule(lr{0.5em}){4-5}
\cmidrule(lr{0.5em}){6-7}
\cmidrule(lr{0.5em}){8-9} 
        & per core & overall & per core & overall & per core & overall & per core & overall \\  
\#cores & Mflop/s  & Tflop/s & Mflop/s  & Tflop/s & Mflop/s  & Tflop/s & Mflop/s  & Tflop/s \\
\addlinespace %%[\smallskipamount]
\D8192 & 539 & 4.4  & 912 & 7.5  & 525 &\D4.3 & 752 &\D6.2  \\
 16384 & --  & --   & --  & --   & 596 &\D9.8 & 783 & 12.8 \\
 32768 & --  & --   & --  & --   & 503 & 16.5 & 771 & 25.3 \\
\bottomrule
\end{tabular}
\end{table}

\subsection{QUDA}

BQCD can run on GPUs by employing the QUDA library \cite{Clark:2009wm}.
QUDA has a BQCD interface to its \emph{cg} and multishift \emph{cg}
solvers.

\section*{Acknowledgements}

We would like to thank Gerrit Schierholz, Roger Horsley, Waseem Kamleh,
Paul Rakow and James Zanotti for support, stimulating discussions and
bug reports.  The computations were performed on a Cray XC40 of the
North-German Supercomputing Alliance (HLRN) and the IBM BlueGene/Q at
Jülich Supercomputer Centre (JSC).

%%%%%%%%%%%%%%%%%%%%%%%%%%%%%%%%%%%%%%%%%%%%%%%%%%%%%%%%%%%%%%%%%%%%%%%%%%%%%
\end{document}